\documentclass[letterpaper,12pt]{article}
\usepackage{graphicx} 
\graphicspath{ {./imagefolder/} }
\usepackage{url}
\usepackage{multicol}
\usepackage{algorithm}
\usepackage[noend]{algpseudocode}
\usepackage{dgjournal}      \usepackage{amsmath}
\usepackage{xcolor}\usepackage{mathptmx}
\usepackage{anyfontsize}
\usepackage{t1enc}

\newtheorem{Lem}{Lemma}
\usepackage{multirow}
\begin{document}

\begin{center}
	
 \textbf{ {\fontsize{20}{60}\selectfont Doubly Robust Adaptive LASSO for Effect Modifier Discovery}}\\

 Asma Bahamyirou $^1$, Mireille E. Schnitzer $^1$, Edward H. Kennedy $^2$, Lucie Blais $^1$, and Yi Yang $^3$
 \\
 
 $1$: Université de Montréal, Faculté de Pharmacie.\\
 $2$: Canergie Mellon University, Department of Statistics \& Data SCience\\
 $3$: McGill University, Department of Mathematics and Statistics.
\end{center}


\begin{abstract}
Effect modification occurs when the effect of the treatment on an outcome differs according to the level of a third variable (the effect modifier, EM). A natural way to assess effect modification is by subgroup analysis or include the interaction terms between the treatment and the covariates in an outcome regression. The latter, however, does not target a parameter of a marginal structural model (MSM) unless a correctly specified outcome model is specified. Our aim is to develop a data-adaptive method to select effect modifying variables in an MSM with a single time point exposure. A two-stage procedure is proposed. First, we estimate the conditional outcome expectation and propensity score and plug these into a doubly robust loss function. Second, we use the adaptive LASSO to select the EMs and estimate MSM coefficients. Post-selection inference is then used to obtain coverage on the selected EMs. Simulations studies are performed in order to verify the performance of the proposed methods.

\end{abstract}
\textbf{ Keywordst}: Doubly robust, Adaptive LASSO, Effect modification, Selective inference.

\section{Introduction}


Effect modification occurs when the effect of a treatment on an outcome differs according to the level of some pre-treatment variables (the effect modifier, EM). Detecting variables that are EMs is not a straight-forward task even for a subject matter expert. A natural way to assess effect modification in experimental and observational studies is to perform subgroup analysis, in which observations are stratified based on the potential EMs after which stratum-specific estimates are calculated, though this becomes infeasible with a greater number of potential effect modifiers. One can also include the interaction terms between the treatment and the potential EMs in an outcome regression analysis. With observational data however, this approach does not target a parameter of a marginal structural model (MSM) unless a correct model for the outcome conditional on confounders, treatments, and EMs is specified. In contrast, MSMs can provide a summary of how effect modification occurs in the absence of confounding. 
 Different methods for the estimation of effect modification have been proposed recently.
 For example, Green and Kern \cite{R1} used Bayesian Additive Regression
Trees (BART) \cite{R2} to model the conditional average treatment effects (CATE). Imai and Ratkovic \cite{R3} studied EM selection by adapting the support vector machine classifier. Nie and Wager \cite{R4} developed a two-step algorithm for heterogeneous treatment effect estimation using the marginal effects and treatment propensities. Lue et al., \cite{R5} used dimension reduction techniques to learn heterogeneity by estimating a lower dimensional linear combination of the covariates that is sufficient to model the regression causal effects. Wager and Athey \cite{R6} proposed a nonparametric approach for estimating heterogeneous treatment effects using a random forest algorithm \cite{R7}. Powers et al., \cite{R8} developed an algorithm for heterogeneous treatment effect estimation by adapting the multivariate adaptive regression splines \cite{R9}.  Zhao et al. \cite{R10} introduced an algorithm based on a semiparametric model that selects the EMs by using Robinson's transformation \cite{R11} and Least Absolute Shrinkage and Selection Operator
(LASSO). Doubly robust semiparametric methods such as Targeted Minimum Loss-Based
Estimation (TMLE) \cite{R12,R13}, which is closely related to previously existing methods \cite{R14,R15}
have been proposed. The term doubly robust comes from the fact that the method
requires both the estimation of the treatment model and the outcome expectation
conditional on treatment and covariates, where only one of which needs to be
correctly modeled to allow for consistent estimation  of the parameter of interest. However, in a situation
where one nuisance parameter is inconsistently estimated, the asymptotic linearity is affected \cite{R16}. Lee et
al. \cite{R17} developed a doubly robust estimator of the CATE along with a uniform confidence band. Rosenblum and van der Laan \cite{R13} developed TMLE for MSMs, which can be used to model effect modification, in non-longitudinal settings. Zheng et al. \cite{R18} developed TMLE for MSMs with counterfactual covariates in longitudinal settings. Most recently, Kennedy \cite{R19} analyzed a version of the pseudo-outcome regression method for CATE estimation and derives model-free error bounds.\\
In this paper as in \cite{R10}, we focus on the selection of pre-treatment EMs in a linear MSM for the CATE with a single treatment time-point. Thus, we consider modifiers of the additive effect
of a treatment on the mean outcome. We use a component of the efficient influence function of the ATE along with the Adaptive LASSO (Zou, 2006) to select EMs. To the best of our knowledge, our paper is one of the first along with \cite{R19,R20} to investigate and apply a doubly robust two-stage regularization for a CATE model. Our estimation approach can be carried out with standard software implementations, is doubly robust (unlike \cite{R10}), can accommodate adaptive methods to estimate the nuisance quantities, and produces estimates of the parameters of an easily interpretable model. A two-stage procedure is thus proposed. First, we estimate two nuisance quantities (the conditional outcome expectation and treatment model) and plug these quantities into a specific function to create a pseudo outcome as developed in \cite{R21,R22,R23}. Second, we take the pseudo outcome and apply the adaptive LASSO \cite{R24} to select the EMs and estimate the MSM coefficients. We then apply post-selection inference in order to produce interpretable confidence intervals after the EM selection by adaptive LASSO. We perform simulation studies in order to verify the performance (selection, estimation, double robustness, and post-selection inference) of the proposed method.\\
The remainder of this article is organized as follows. In Section 2, we use the potential outcomes framework to define the target causal
parameter of interest and describe our proposed estimation approach. In Section 3, we conduct a simulation study to verify the performance (selection, MSM coefficient estimation, and double robustness) of the proposed method in both low and high dimensional settings. We present an analysis of the safety of asthma medications
during pregnancy in Section 4. A discussion is provided in Section 5.

\section{Methods}
\label{sec2}

In this section, we present our development of the methodology for the selection of the EMs.
\subsection{The framework}

The observed data, $\{(\boldsymbol W_i,A_i,Y_i) \}^{n}_{i=1}$, are comprised of independent and identically distributed samples of $O=(\boldsymbol W,A,Y)  \sim P_0$, where $\boldsymbol W$ is the baseline covariates of a patient, $A$ is the binary treatment which equals $1$ if the patient received treatment and $0$ otherwise, and $Y$ is the observed outcome (binary or continuous). Let $\boldsymbol V$ represent the subset of the variables in $\boldsymbol W$ that represents the potential EMs of interest. We use $O_i=(\boldsymbol W_i,A_i,Y_i)$ to represent the i-th observation of the data. In order to define the target parameter, we use the counterfactual framework of Rubin \cite{R25}.
 Let $Y^a$ denote the potential (or counterfactual) outcome that would have occurred under the treatment value $A = a$.  In this paper, we focus on marginal models for the CATE. If we assume that we observe $Y=Y^a$ when $A=a$ (consistency \cite{R26}, no interference, positivity and no unmeasured confounders \cite{R27}), the CATE can be defined and identified nonparametrically as:
\\
\begin{equation}
\begin{array}{lll}
\psi_0(\boldsymbol V) & = & E_0\{  Y^1-Y^0 |\boldsymbol V \} \\
       & = & E_{\boldsymbol W|\boldsymbol V}\{ \underbrace{E_0(Y|A=1,\boldsymbol W)}_{\bar{Q}_0(1,\boldsymbol W)}-\underbrace{E_0(Y|A=0,\boldsymbol W)}_{\bar{Q}_0(0,\boldsymbol W)} )|\boldsymbol V\} \\
       & = &  E_{W|\boldsymbol V}\{ \bar{Q}_0(1,\boldsymbol W)-\bar{Q}_0(0,\boldsymbol W)|\boldsymbol V\}
\end{array} 
\end{equation}
where $E_0$ is the expectation with respect to the outcome and $E_{W|V}$ is the
expectation conditional on the baseline covariates. In this work, we choose to model the CATE using a linear regression model defined as $\tilde{\psi}_0(\boldsymbol V) = \beta_0+\boldsymbol V^T\boldsymbol \beta_V$ where the relevant subset of $\boldsymbol V$ will be selected using adaptive LASSO \cite{R24}. Our goal here is to identify the true EMs among the set $\boldsymbol V$, and estimate their associated coefficients. One could use non-linear models or machine learning methods to estimate $\tilde{\psi}_0(\boldsymbol V)$, which is important when the goal is
prediction \cite{R37} (e.g. for personalized medicine). However, if interpretation of the coefficient associated with each $V^{(s)}$ is important, it may be beneficial to use a linear model rather than a black box approach \cite{R28}. 

\subsection{Adaptive LASSO}\label{sec_predLASSO}
The adaptive LASSO \cite{R24} is an extension of the traditional LASSO of Tibshirani \cite{R29} that uses
coefficient specific weights. Zou \cite{R24} showed that the
adaptive LASSO estimator has the oracle property which roughly means that the algorithm identifies the right subset of variables (consistency of variable selection) and that the coefficient estimators of the selected variables are asymptotically normal. In a prediction (non-causal) setting, let $Y$ be an observed outcome and $\boldsymbol V$ a set of covariates. Under the linear model, we can select predictors of Y by solving the equation below:
\begin{equation}
 \arg \min_{\alpha',\boldsymbol\beta'}  \sum_{i=1}^{n}( Y-\alpha{'}-\boldsymbol V_i^T\boldsymbol\beta')^2+\lambda\sum_{j=1}^{p}\widehat{w}_j |\beta_j'|
\end{equation}
where $\boldsymbol\beta' = (\beta_1',...,\beta_p')$, $\widehat{w}_j = 1 / |\tilde{\beta}_j'|^{\gamma}$, for some $\gamma > 0 $ and $\tilde{\beta}_j'$ is a $\sqrt{n}$-consistent estimator of $\beta_j'$. The selected variables are the positions of the non-zero entries of the solution of (2).
When the sample size grows, the weights associated with the zero-coefficient predictors tend to infinity, while the weights corresponding to true predictors converge to a constant. Thus, true-zero coefficients are less likely to be selected by the adaptive LASSO than by the standard LASSO, which does not have the oracle property \cite{R24}.
\subsection{Highly Adaptive LASSO (HAL)}

Assume $E(Y|V)$ a regression function where $Y$ is the observed outcome and $V$ is the set of covariates. Consider a map of $\boldsymbol V$ onto a set of binary indicator basis functions. For example, if $\boldsymbol V$ is scalar, we generate for an observation $v$, $\boldsymbol \phi^*(v)=(\phi_1^*(v),...,\phi_n^*(v))^T$, where $\phi_i^*(v)=I(v\geq V_i)$, for $i=1,...,n$. With two dimensions, $\boldsymbol V=(V^{(1)},V^{(2)})^T$, we need to include the second order basis functions $\boldsymbol \phi_i^*(\boldsymbol v)=I(v_1\geq V^{(1)}_i,v_2\geq V^{(2)}_i)$, for $i=1,...,n$. The HAL estimator \cite{R30} is obtained by fitting a $L_1$-penalized regression of the outcome $Y$ on these basis functions, with the
optimal $L_1$‐norm chosen via cross‐validation. The HAL estimator of the regression function $E(Y|\boldsymbol  V)$  converges to the true regression function  in $L_2$-norm no slower than $n^{-1/4}$ regardless of the dimension of $\boldsymbol V$, under the assumption that the regression function has bounded variation norm. 
\subsection{Selective inference}

Let $ \widehat{\boldsymbol \beta}'$ be the solution of (2) and $\widehat{\boldsymbol \beta}_{\widehat{M}}'$ the non-zero subvector of $\widehat{\boldsymbol \beta}'$ where $\widehat{M} \subseteq \{1, ..., p\}$ corresponds to the positions of the non-zero entries. Suppose that we are interested in making inference for $\widehat{\boldsymbol \beta}_{\widehat{M}}'$ in the prediction model of Section~\ref{sec_predLASSO}.  A naive way to obtain inference after selecting the covariates in the model is the standard hypothesis tests for linear regression that treat $M$, representing the non-zero entries of $\boldsymbol \beta'$ and thus the true model, as known. It is easy to see that $\widehat{\boldsymbol \beta}'$ depends on the selected model $\widehat{M}$. Therefore, Lee et al., \cite{R31} studied the conditional distribution $\widehat{\boldsymbol \beta}_{M}'|\{\widehat{M}=M\}$ and showed that this conditional distribution is a truncated normal Gaussian. They constructed a pivotal statistic for $\widehat{\boldsymbol \beta}_{\widehat{M}}'$ which can be used for hypothesis testing and therefore by test inversion, to construct a confidence interval. Let $F(y;\mu,\sigma^2,l,u)$ be the CDF of a normal $N(\mu,\sigma^2)$ truncated to the interval $[l,u]$, $e_j$ the unit vector for the j-th coordinate so that $(\widehat{\beta}_{M}')_j=\eta^T_{M}Y$,  $\eta_{M} =[(\boldsymbol V^T_{M} \boldsymbol V_{M})^{-1} \boldsymbol V^T_{M}]^Te_j$ and $\sigma^2_{*} = \sigma^2\eta^T_{M}\eta_{M}$. In the linear regression setting where $Y\sim N(\mu,\sigma^2I_n)$,  Lee et al., \cite{R31} showed that  $F((\widehat{\beta}_{M}')_j;(\beta_{M}')_j,\sigma^2_{*},\nu^-,\nu^+)|\{\widehat{M}=M\} \sim Unif(0,1)$, where $[\nu^-,\nu^+]$ is defined in \cite{R31} as a function of $Y$ and the model $M$.  By inverting the hypothesis testing, we can find a $(1-\alpha)$ confidence interval for $(\widehat{\beta}_{M}')_j$, conditional on $\widehat{M}=M$, by finding $[L^*,U^*]$ such that
 $$F((\widehat{\beta}_{\widehat{M}}')_j;L^*,\widehat{\sigma}^2_{*},\nu^-,\nu^+)|\{\widehat{M}=M\}= 1-\alpha/2$$
 and 
  $$F((\widehat{\beta}_{\widehat{M}}')_j;U^*,\widehat{\sigma}^2_{*},\nu^-,\nu^+)|\{\widehat{M}=M\}= \alpha/2$$
  
 In this next section, we will explain how this result is applied in our setting.

\subsection{The model}

\subsubsection{Model definition}
Let $\psi_0(\boldsymbol V)= E_0\{  Y^1-Y^0 | \boldsymbol V \}$ be the CATE. Denote $\bar{Q}_0(a,\boldsymbol W) = E_0(Y|A=a,\boldsymbol W)$, the outcome expectation, and $g_0(a|\boldsymbol W)=P(A=a|\boldsymbol W)$ as the propensity score. We suggest to use the doubly robust and efficient loss-function
proposed by van der Laan \cite{R21}, inspired by Rubin and van der Laan
\cite{R32},
$ 
L_{Q_0,g_0}(\psi)(O)=( D(\bar{Q}_0,g_0)(O)- \psi_0(\boldsymbol V) )^2
$ 
where 
\begin{equation}
D(\bar{Q}_0,g_0)(O)=\displaystyle\frac{2A-1}{g_0(A|\boldsymbol W)}(Y-\bar{Q}_0(A,\boldsymbol W))+\bar{Q}_0(1,\boldsymbol W)-\bar{Q}_0(0,\boldsymbol W)
\end{equation}
is indexed by the nuisance parameters $(\bar{Q}_0; g_0)$. A similar pseudo-outcome is also
used in Zhao et al. \cite{R22} for estimating optimal individualized treatment rules and
Kennedy et al. \cite{R23} for the estimation of continuous treatment effects. \\
The next lemma shows that if one of the two nuisance quantities are consistent, the CATE can be obtained by the conditional expectation of the estimated pseudo-outcome.
\begin{Lem}
Let $||f||_{2,P_0}^2 = \int f(z)^2 dP_0(z) $ denote the $L2(P)$ norm. Suppose either $\bar{Q}_n$  converges to $\bar{Q}_0$ or $g_n$ converges to $g_0$ in the sense that $E||\bar{Q}_n - \bar{Q}_0||^2 = o(1)$ or $E||g_n - g_0||^2 =o(1)$ (not necessarily both). Then $E(D(\bar{Q}_n,g_n)(O) |\boldsymbol V) \rightarrow \psi_0(\boldsymbol V)$ as $n \rightarrow \infty$.
\end{Lem}
The preceding lemma shows that the pseudo-outcome we propose for the CATE is doubly-robust in the sense that if at least one nuisance
estimator ($\bar{Q}_n$ or $g_n$) converges to the correct function, but not necessarily both, then a regression of the pseudo-outcome onto the effect modifiers will be consistent for the CATE. Adding and subtracting the true CATE is the key idea to prove Lemma 1. Then, the regression function of the pseudo-outcome on $V$ can  be split into two terms: the true CATE and a second term that is a function of both $\bar{Q}_n - \bar{Q}_0$ and $g_n - g_0$. See the Appendix for the proof of Lemma 1.

Suppose that an investigator would like to identify the true EMs amongst multiple suspected effect modifying variables $\boldsymbol V=(V^{(1)},...,V^{(p)})$. As described above, to accomplish this we use a linear model for the CATE with corresponding MSM defined as  $\tilde{\psi}_0(\boldsymbol V) = \beta_0+\boldsymbol V^T\boldsymbol\beta_V$ under a least squared error loss function. We then use the adaptive LASSO estimator \cite{R24} to select amongst the $V^{(j)}$s. More specifically, as suggested by Rubin and van der Laan \cite{R33}, we penalize the aforementioned loss function $L_{\bar{Q}_0,g_0}$ by the adaptive LASSO penalty. Let $D_n=D(\bar{Q}_n,g_n)(O)$ be the estimated pseudo outcome. The parameters of the MSM $\boldsymbol\beta=(\beta_0,\beta_1, ... ,\beta_p)$ are estimated by minimizing the risk function below:
\begin{equation}
\widehat{\boldsymbol \beta} = \arg \min_{\beta}  \sum_{i=1}^{n}( D_{i,n}-\tilde{\psi}_0(\boldsymbol  V_i))^2+\lambda\sum_{j=1}^{p}\widehat{w}_j |\beta_j|
\end{equation}
where $\widehat{\omega}_j = 1 / |\tilde{\beta}_j|^{\gamma}$, for some $\gamma > 0 $ and $\tilde{\beta}_j$ is a $\sqrt{n}$-consistent estimator of $\beta_j$.\\
An optimal method would possess the oracle property, able to select the appropriate variables and unbiasedly estimate the selected parameters. 
Let $\mathbf{A}$  be the set of true variables in the model and $\mathbf{A}_n^*$ be the set selected using adaptive LASSO.
\begin{Lem}
Let $D=D(Q_0,g_0)(O)$ be the oracle pseudo-outcome, depending on the true outcome expectation and propensity score. Assume $E(D|\boldsymbol V) = \beta_0+\boldsymbol V^T\boldsymbol\beta_V$ and $|\mathbf{A}| = p_0 < p$. Suppose that $\lambda/\sqrt{n}\rightarrow 0$ and $\lambda n^{(\gamma-1)/2} \rightarrow \infty$.
The proposed estimator $\widehat{\boldsymbol \beta}$ inherits the adaptive LASSO oracle properties, i.e. 
\begin{itemize}
    \item Consistency in variable selection (i.e. identifies the right subset model):\\ $\lim_{n \rightarrow \infty} P( \mathbf{A}_n^*= \mathbf{A}) = 1$.
    \item Asymptotic normality (i.e. has the optimal estimation rate): $\sqrt{n}(\widehat{\boldsymbol \beta}_{\mathbf{A}}-\boldsymbol \beta_{\mathbf{A}})  \rightarrow_{d} N(0,\Sigma^*)$, where $\Sigma^*$ is the covariance matrix knowing the true subset model and $\widehat{\boldsymbol \beta}_{\mathbf{A}}$ is the coefficient estimates resulting from the Adaptive LASSO regression of $D$ on $V$.
\end{itemize}

\end{Lem}
As a consequence, our proposed estimator is able to select the correct subset of EMs and produce an unbiased estimate of the MSM coefficients in large samples. See the Appendix for the proof of Lemma 2.

\subsubsection{Estimation}
In this paragraph, we describe how our proposal can be easily implemented in a two-stage procedure. In the first stage, we construct the pseudo-outcome function by producing estimates $\bar{Q}_n(a,\boldsymbol W)$ and $g_n(a|\boldsymbol W)$ of the two nuisance quantities  and plugging them into $D$. Machine Learning (ML) methods are often recommended \cite{R13} for estimating $\bar{Q}_n$ and $g_n$. In the second stage, we run the adaptive LASSO regression of the estimated pseudo-outcome $D(\bar{Q}_n,g_n)(O)$ on the set $\boldsymbol V$. The selected EMs correspond to the non-zero coefficients of the adaptive LASSO regression.\\
The proposed algorithm for estimating the parameters in the CATE model with a given value of $\lambda$ is as follows: 

\begin{algorithm}[H]
\caption{Effect modifiers adaptive LASSO algorithm }\label{alg:euclid}
\begin{algorithmic}[1]
\State Estimate the outcome expectation $\bar{Q}_n(a,\boldsymbol W)=\hat{E}(Y|A=a,\boldsymbol W)$ for each subject.
\State Obtain the estimated propensity score $g_n(a|\boldsymbol W)=\hat{P}(A=a|\boldsymbol W)$ for each subject.
\State Construct an estimate of the doubly robust function $D_n$ by plugging in the estimated $\bar{Q}_n$ and $g_n$.
\State Select the effect modifiers by following steps (a)-(d) below:
\begin{enumerate}
\item[(a)] Run a linear regression of $D_n$ on $\boldsymbol V$ as the set of covariates. Obtain $\tilde{\beta}_j$, the estimated coefficient of $V^{(j)}$, $j=1,...,p$.
\item[(b)]  Define the weights $\widehat{\omega}_j = \frac{1}{ |\tilde{\beta}_j|^{\gamma}}$, $j=1,...,p$ for some $\gamma > 0 $.
\item[(c)] Run a LASSO regression of  $D_n$  on $\boldsymbol V$ with $\widehat{\omega}_j$ as the penalty factor associated with $V^{(j)}$ with a given $\lambda$. 
\item[(d)] The non-zero coefficients of the solution of the adaptive LASSO regression $\{\widehat{\beta}_j\}^{p}_{j=1}$ are the selected effect modifiers. 
\end{enumerate}
\State The final estimate of the CATE is  $\psi_n(\boldsymbol V) = \widehat{\beta}_0+ \sum_{j=1}^{p} V^{(j)}\widehat{\beta}_j$.
\end{algorithmic}
\end{algorithm}

For the adaptive LASSO tuning parameters, we choose $\gamma=1$ (Nonnegative Garotte Problem \cite{R34}) and $\lambda$ is selected using cross-validation as suggested by Zou \cite{R24}. The traditional cross-validation minimizes the prediction error knowing the true outcome. In our setting, the Adaptive LASSO is run with the estimated pseudo-outcome as the ``true" outcome. We conjecture that if the two nuisance parameters are consistently estimated at fast enough rates, we should be able to use the estimated pseudo-outcome to find an optimal tuning parameter. This conjecture agrees with recent results from Kennedy (2020) \cite{R19}. Naive inference by ignoring the EM selection would result in incorrect confidence intervals. Zhao et al. \cite{R10} showed that when the outcome is observed with error, the selective pivotal statistic proposed by Lee et al. \cite{R31} is still asymptotically valid. Thus we apply their methodology which is expected to produce valid asymptotic results as long as $\bar{Q}_n$ is consistent and both $\bar{Q}_n$ and $g_n$ converge faster than at a $n^{1/4}$ rate in the $l_2$ norm \cite{R35}. In order to construct a selective $95\%$-confidence intervals for the selected submodel, we use the R package \textbf{selectiveInference} \cite{R36} for post-selection inference. The estimated $\widehat{\sigma}^2$ used in the package is the variance of the residual from fitting the full model in $4(a)$.

\section{Simulation study}
\label{sec3}

\subsection{Data generation and parameter estimation}

To evaluate the performance of the proposed method in finite samples, we conducted a simulation study under four scenarios. We simulated data $O = (\boldsymbol W,A, Y)$ representing baseline
covariates $\boldsymbol W$, a binary exposure $A$, and a continuous outcome $Y$. The baseline covariates $\boldsymbol W$ include three confounders $(X, V^{(1)}, V^{(2)})$, one instrument $Z$ (pure cause of treatment), and two pure causes of the outcome $(V^{(3)}, V^{(4)})$. All covariates were generated independently with the Bernoulli distribution with success probability $p$: $X \sim B(p=0.4)$, $V^{(1)} \sim B(p=0.5)$, $V^{(2)} \sim B(p=0.6)$, $V^{(3)} \sim B(p=0.5)$, $V^{(4)} \sim B(p=0.7)$ and $Z \sim B(p=0.45)$. \\
We varied the strength of the relationship between covariates, outcome and treatment across three low-dimensional scenarios. In the first, we used an outcome model where the covariates were strongly predictive, and a treatment model where the covariates were weakly predictive. The treatment mechanism $g_0$ was set as a Bernoulli with the probability generated linearly in the three confounder variables and single instrument,
$$P_0(A=1|X)=\textrm{expit}\{ 0.5Z -0.2X +0.3V^{(1)}1+ 0.4V^{(2)}\}$$ 
where $\textrm{expit}(x)=1/\{ 1+exp(-x)\}$. The observed continuous outcome $Y$ was linearly generated as:
 $$Y = 1+ A  -0.5X +2V^{(1)} +V^{(2)}  +V^{(3)} -0.2V^{(4)} + 4V^{(1)}V^{(2)}V^{(3)} +A(0.5V^{(1)}+V^{(3)})  +N(0,1)$$ 
The effect modification arises due to interaction between treatment and covariates.\\ The second scenario has the same data generation except that the coefficient of the interaction term $V^{(1)}V^{(2)}V^{(3)}$ is $0$ instead of $4$. In the third scenario, we use an outcome model where the covariates are weakly predictive, and a treatment model where the covariates are strongly predictive. We focus here on the first scenario and describe all other simulations settings and results in the Appendix.\\ We thus have two EMs $(V^{(1)},V^{(3)})$, where the first is a confounder and the second is a pure cause of the outcome. In practice, we are not aware of the true data generating mechanism. So we have a potential set of EMs: $\boldsymbol V=(V^{(1)},V^{(2)},V^{(3)},V^{(4)})$.                                                  Let $\psi_0(\boldsymbol V)=E_{P_0}(Y^1-Y^0|\boldsymbol V)$ be the true (nonparametric) CATE, which we model as an MSM: $\tilde{\psi}_0(\boldsymbol V)=\beta_0+\beta_1V^{(1)}+\beta_2 V^{(2)} +\beta_3 V^{(3)} + \beta_4 V^{(4)}$. Our goal here is to identify among the set $\boldsymbol V$, the true EMs and estimate their associated coefficients. Given the data generated, the true values of the coefficients are $\boldsymbol\beta_v=(0.5,0,1, 0)$. We set $n=1000$ and then $10000$. We also add a smaller sample size $n=100$ with results in the appendix.\\ To evaluate the performance of our method in high-dimensional settings, we also extend the first scenario by adding $50$ pure binary noise covariates (unrelated to treatment or outcome) to our set of covariates, which are included as potential confounders and EMs.  The true values of the coefficients in the MSM are thus $\boldsymbol\beta_v=(0.5,0,1, 0,...,0)$.\\  
Under each low-dimensional scenario, we tested our proposed method under four different implementations:
\begin{enumerate}
\item[ (1)] Qcgc: Both of the models for $\bar{Q}$ and $g$ are correctly specified using generalized linear models (GLMs).
\item[ (2)]Qc: Only the GLM for $\bar{Q}$ is correctly specified. $g$ is misspecified using a logistic regression of treatment $A$ on variable $X$.
\item[ (3)]gc:  Only the GLM for $g$ is correctly specified. $\bar{Q}$ is misspecified using a GLM of treatment $Y$ on variables $A$ and $V^{(3)}$.
\item[ (4)]HAL: Both $\bar{Q}$ and $g$ are estimated using the Highly Adaptive LASSO (HAL) \cite{R30,R37}. We use the package default setting.
\end{enumerate}
For comparison, we also tested two implementations of a linear regression model for the outcome to directly assess effect modification:
\begin{enumerate}
    \item[ (5)]NLin:  Linear regression with main terms (treatment and all covariates) and interactions between treatment and covariates. Only first-order interactions were included.
    \item[ (6)]CLin: Linear regression with a correctly specified outcome model.
\end{enumerate}
 Standard confidence intervals are presented for the linear model case and, in our summary, a p-value of less than $0.05$ is used as a criterion for a variable to be selected. 
 In the higher dimensional scenario, only HAL was used to estimate $\bar{Q}$ and $g$.
 
\subsection{Simulation results}
For each scenario, we produced boxplots of the MSM coefficient estimates. We also present the percent selection, the coverage proportion of the confidence intervals and the false coverage rate in order to summarize the average performance of each estimator and implementation. The percent selection for our LASSO method was obtained as the percentage of estimated coefficients that are non-zero throughout the 1000
generated datasets, and for the linear regression, the percentage of p-values $<0.05$. The coverage for each true effect modifier was obtained as the number of times the true model was selected and the corresponding confidence intervals contained the true coefficients, divided by the number of times the true model was selected. For the linear regression, the percent coverage was instead calculated for each coefficient and defined as the proportion of the confidence intervals that contained the true coefficient throughout the 1000 generated datasets. The false coverage rate (FCR) for our LASSO model was obtained as the number of non-covering confidence intervals among the selected coefficients, divided by the number of the selected coefficients throughout the 1000
generated datasets \cite{R31}.  

\begin{figure}[H]
\includegraphics[width=15cm]{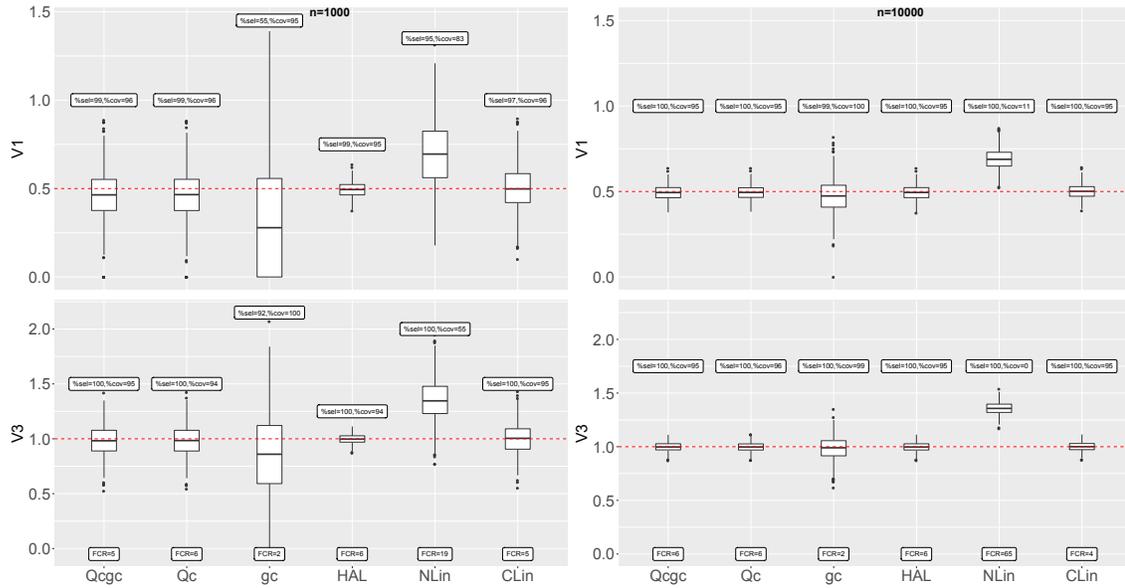}
\caption{Simulation results illustrations (Data generating scenario 1). Box plots of $1000$ MSM coefficients estimates for the true EMs $(V^{(1)},V^{3)})$. The true values of the coefficients are $(0.5, 1)$. Notation: Qcgc: models for $\bar{Q}$ and $g$ are correctly specified, Qc: $\bar{Q}$ is correctly specified, gc: $g$ is correctly specified, HAL: $\bar{Q}$ and $g$ are estimated with HAL, NLin: Naive linear model, CLin: Correct linear model. $\%sel$: percent selection of a covariate $\times 100$, $\%cov$: coverage rate of the confidence interval of a coefficient estimate $\times 100$, $FCR$: False coverage rate of the model $\times 100$.}
\label{fig:sim1}
\end{figure}

\begin{figure}[H]
\includegraphics[width=1.2\textwidth]{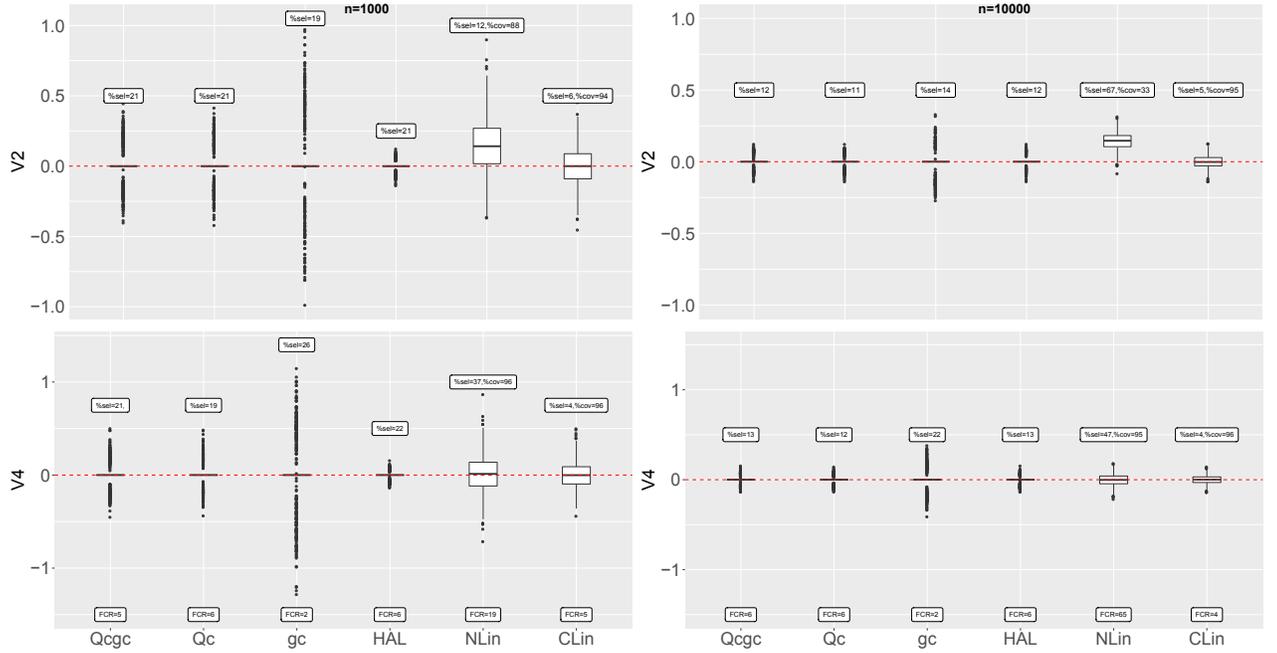}
\caption{Simulation results illustrations (Data generating scenario 1). Box plots of $1000$ MSM coefficients estimates for the non-EMs $(V^{(2)},V^{4)})$. The true values of the coefficients are $(0, 0)$. Notation: Qcgc: models for $\bar{Q}$ and $g$ are correctly specified, Qc: $\bar{Q}$ is correctly specified, gc: $g$ is correctly specified, HAL: $\bar{Q}$ and $g$ are estimated with HAL, NLin: Naive linear model, CLin: Correct linear model. $\%sel$: percent selection of a covariate $\times 100$, $\%cov$: coverage rate of the confidence interval of a coefficient estimate $\times 100$, $FCR$: False coverage rate of the model $\times 100$.}
\label{fig:sim1suite}
\end{figure}

For the first low-dimensional scenario, Figures \ref{fig:sim1} and \ref{fig:sim1suite} contain the boxplots of the MSM coefficient estimates for the true EMs $(V^{(1)},V^{3)})$ and non-EMs $(V^{(2)},V^{(4)}$), respectively. Table \ref{table:simu1} (in the Appendix) contains the numerical results. As shown in the first two boxplots in Figures \ref{fig:sim1} and \ref{fig:sim1suite}, the implementations (1) Qcgc and (2) Qc performed very well. We obtained unbiased estimates and a coverage of the confidence interval that tended to be around $95\%$. The FCR was close to the optimal 0.05. In the third boxplot, corresponding to  implementation (3) gc, where only the propensity score was correctly specified, the estimator was more biased for both sample sizes but had higher coverage rates and lower FCR. In the fourth boxplot where the estimator was implemented with HAL, the estimator performed well across all measures. In all implementations the true effect-modifiers $(V^{(1)},V^{(3)})$  were selected around 100 percent of the time except when only the propensity score was correctly specified for the smaller sample size (gc). The percent selection of variables that are not effect-modifiers $(V^{(2)},V^{(4)})$ was around $20\%$ for $n=1000$. In implementations (1), (2), and (4), the percentage was almost halved for $n=10000$. The FCR was controlled around the nominal $0.05$ level in all situations even when only one nuisance model was correctly specified. This supports the double robustness of the proposed estimator and the appropriateness of the post-selection confidence intervals. In implementation (5) NLin, the naive linear model with a misspecified term performed poorly, even when increasing the sample size. On the other hand, when the linear model was correctly specified in implementation (6) CLin, the  coefficient estimates were unbiased on average and the coverage was near-optimal. For the two other data generating scenarios described at more length in the Appendix, the results (Tables \ref{table:simu2} and  \ref{table:simu3}) look similar to those in the first scenario.\\

\begin{figure}[H]
\includegraphics[width=15cm]{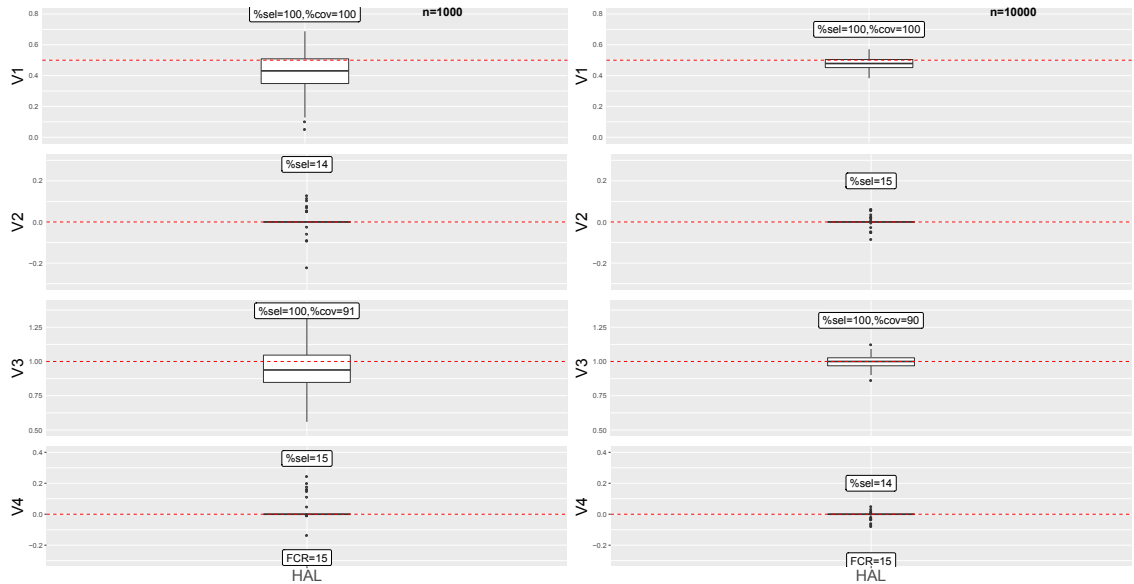}
\caption{Simulation results for high-dimensional setting (Data generating scenario 1). Box plots of MSM coefficients estimates over 100 simulations for the potentials EMs $\boldsymbol V=(V^{(1)},V^{(2)},V^{(3)},V^{(4)})$. The true values of the coefficients are $(0.5, 0, 1, 0)$. Notations: HAL: $\bar{Q}$ and $g$ are estimated with HAL, $\%sel$: percent selection $\times 100$, $\%cov$: coverage rate $\times 100$, $FCR$: False coverage rate $\times 100$.}
\label{fig:simHD}
\end{figure}

\begin{figure}[H]
\includegraphics[width=14cm]{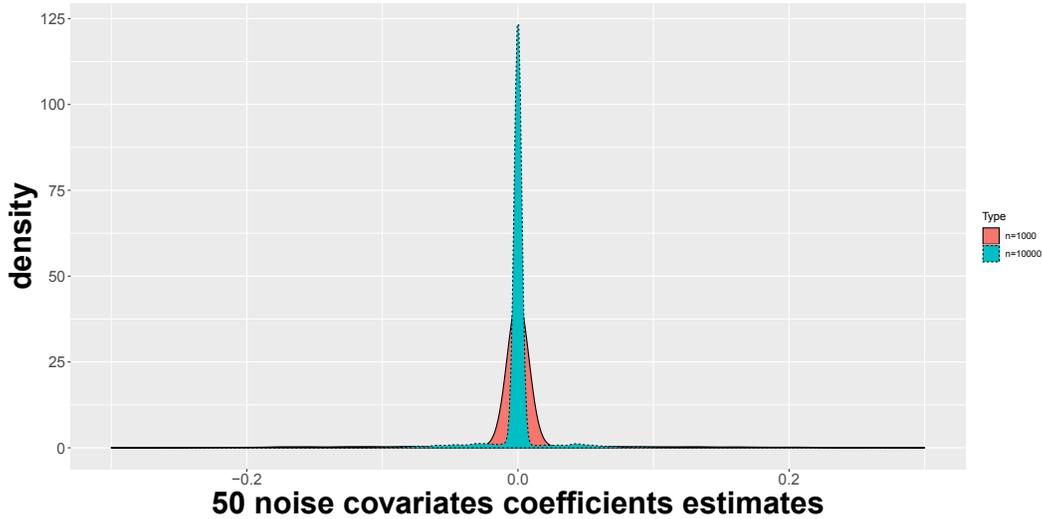}
\caption{Illustrations for high-dimensional setting. Box plots of the MSM coefficients estimates over 100 simulations for the $50$ noise covariates (for both $n=1000$ and $n=10000$). The true values of the coefficients are $(0,...,0)$.}
\label{fig:simNoise}
\end{figure}

 Table \ref{table:smallsize} in the Appendix contains the results with the small sample size $n=100$. The performance of the proposed methods decreased across all measures except for $V^{(3)}$ where there was a higher coverage rate when $\bar{Q}$ and $g$  were correctly specified or estimated with HAL \\The results of the high-dimensional setting are presented in Figures \ref{fig:simHD} and \ref{fig:simNoise}. $\bar{Q}$ and $g$  were estimated with HAL. The estimates were taken over 100 generated datasets and look similar to Figures \ref{fig:sim1} and \ref{fig:sim1suite} for the covariates $V=(V^{(1)},V^{(2)},V^{(3)},V^{(4)})$ in common. For the noise covariate coefficients, the estimates, given in the density plot of Figure \ref{fig:simNoise}, were unbiased for 0. The noise covariates had a low percent selection (see Table \ref{table:HD}). Using median statistics, the noise covariates were selected  around $14\%$ of the time and that proportion decreased to $13\%$ as we increased the sample size. The FCR exceeded the nominal $5\%$ level and was around $15\%$. 
 
In summary, Table \ref{table:simu1} demonstrates that in low-dimensional settings, the proposed algorithm is able to produce unbiased estimates and control the FCR around the nominal level. In contrast, Table \ref{table:HD} demonstrates that in the context of high-dimensional covariates with many candidate EMs, the FCR is generally much larger than the nominal level. Similar results were obtained by Zhao et al. (\cite{R10}, Figure 2). In addition, at least some non-EMs were always selected by the algorithm at the sample sizes investigated.

\section{Data analysis: Asthma medication during pregnancy}

\subsection{Data}

Our data were obtained from a cohort (Firoozi et al. \cite{R38}) of deliveries of  pregnant women with asthma in order to study the effect of using inhaled corticosteroids (ICS) during pregnancy on birth weight. The population of interest is pregnant women with mild asthma and a singleton delivery in Qu\'ebec, Canada between $1998$-$2008$, aged $\leq 45$ years. 
For simplicity, We considered only the first delivery for each woman in this period. Asthma severity was defined according to an index that is based on the Canadian Asthma Consensus Guidelines (Cossette et al. \cite{R39}). A total of $4,707$ pregnancies
in our database fell into this category. ICS exposure was classified in two categories: \textquotedblleft use\textquotedblright (a woman who filled at least one prescription of ICS during pregnancy) and \textquotedblleft no use\textquotedblright (a woman who did not fill any prescription of ICS during pregnancy). The outcome of interest is birth weight (continuous in kilograms). We identified a variety of maternal baseline variables. These potential confounders measured in the year before pregnancy include demographic characteristics (e.g. income security provider and place of residence), chronic diseases (e.g. hypertension and diabetes) and variables related to asthma (e.g. at least one hospitalization for asthma, at least one emergency department visit for asthma, and oral corticosteroids). We also included the cumulative daily dose of ICS in the year before pregnancy and sex of the newborn as potential confounders. A full list of measured potential confounders can be found in Table \ref{table:Baseline} in the Appendix. As we do not know which variables are effect modifiers, we included a wide range of variables in the set $\boldsymbol V$, 22 variables in all. Specifically, these variables were: In the year before pregnancy: at least one dose of inhaled short-acting $\beta_2$-agonists (SABA) taken per week,  medication for epilepsy, use of warfarin, use of beta blockers, asthma exacerbation, oral SABA use, oral corticosteroids, leukoteriene-receptor antagonists, intranasal corticosteroids, at least one hospitalization for asthma, at least one emergency department visit for asthma, and welfare recipient; At the start of the pregnancy: chronic obstructive disease, cyanotic heart disease, obesity, uterine disorder, antiphospholipid syndrome, sex of the newborn, rural/non-rural residence indicator, hypertension, diabetes, and chromosomal anomalies. 

For our pregnancy cohort, the average treatment effect is the expected difference in the mean counterfactual birth weight if all women were exposed to ICS during pregnancy versus the counterfactual birth
weight if all women were not \cite{R40}. The target parameters are the coefficients $\beta_j$, $j=1,...,22$ of the MSM defined as: $\tilde{\psi}_0(\boldsymbol V)  = \beta_0+\sum_{j=1}^{22} V^{(j)} \beta_j$, with $\boldsymbol V=(V^{(1)},..,V^{({22})})$ the set of potential EMs. Taking the sex of the newborn as an EM for example ($V^{(j)} = sex $), $\beta_{j}$ is the difference in the CATE for women having male vs female children.

\subsection{Results}
Baseline characteristics of the pregnancy cohort are presented in Table \ref{table:Baseline}. We first implemented a standard linear regression with main terms for all potential confounders and interaction terms between the treatment and the set $\boldsymbol V$. The estimates of the coefficients of the interaction terms are given in Table \ref{table:LinearAppli}. A variable was considered to be selected as an EM in the standard linear regression if the coefficient of the interaction term between that variable and the treatment had a p-value $< 0.05$. This model concluded that leukoteriene-receptor antagonists and chromosomal anomalies are EMs. In addition, we implemented our LASSO methods using HAL for the estimation of the outcome expectation and propensity score. All of the
covariates were included in the propensity score model as well as in the outcome model. Due to larger weights, a $5\%$ truncation for the values of $g_n$ was used. The selected coefficients of the MSM and their estimated values are presented in Table \ref{table:MSMappli}. Three covariates (leukoteriene-receptor antagonists, warfarin one year before pregnancy, and chromosomal anomalies) were selected using the adaptive LASSO and two of them were significant (leukoteriene-receptor antagonists and chromosomal anomalies) using post-selection inference. Leukoteriene-receptor antagonists and chromosomal anomalies were thus selected as EMs in the association of taking ICS during pregnancy on birth weight. Although the naive linear model and our algorithm generate
very similar sets of EMs, the coefficients of the selected EMs are different (compare Table \ref{table:LinearAppli} with Table \ref{table:MSMappli}). For example, the estimated coefficient of leukoteriene-receptor antagonist is around $-0.17$ in the adaptive LASSO while it is $-0.365$ using the linear model.

\section{Discussion}

In this paper, we proposed a doubly robust estimator for selecting effect modifiers (EMs) in an MSM for the CATE. We used the post selection inference method of Lee et al. \cite{R31} to produce post-selection confidence intervals. Through simulation studies, we studied the performance of the proposed estimator. As well, we showed that our proposed estimator is doubly robust and performs well in a high dimensional setting but had a higher FCR along with an over-selection of non-EMs. We observed a slower convergence of our estimator when the outcome expectation model was misspecified. We also illustrated that the post-selection confidence interval produces good coverage proportions for the selected EMs. In a high dimensional case, we confirmed the observation of Zhao et al. \cite{R10} concerning the FCR which exceeded the nominal level in the presence of many noise covariates. Debiased Lasso \cite{R41} could be considered here in a high dimensional case as proposed in Zhao et al (2017).  In general, the overall performance of our estimator improved with the sample size. However,
the blind usage of traditional methods like a regression with main terms and interactions between treatment and potential effect modifiers may produce biased results. We also show theoretically that our estimator is doubly robust and also inherits the oracle properties of the adaptive LASSO. \\
In our application, the results suggest that leukoteriene-receptor antagonists and chromosomal anomalies may modify the effect of ICS during pregnancy on birth weight for women with mild asthma. The estimated CATE is 0.18 lower for women taking leukoteriene-receptor antagonists. As leukoteriene-receptor antagonists are an addition to ICS, we can suppose that it is a marker for more severe asthma. In the presence of a chromosomal anomaly, the effect of ICS was estimated to be 0.78 lower.  The linear regression with standard significance testing suggested the same but with different coefficient estimates. Such discrepancy may be due to the fact that the naive model doesn't target MSM parameters and thus may not be able to model effect modification in the absence of confounding. The proposed method is doubly robust and can control the FCR in a low dimensional setting. However, in this finite sample setting, it may possibly have shrunk the coefficient values relative to the truth since it is a regularization method. Our results point to the importance of using robust methodologies for selecting effect modifiers in well-defined causal models for estimating the conditional treatment effect.

\section*{Acknowledgments}
This work was supported by the Natural Sciences and Engineering Research Council of Canada
(Discovery Grant and Accelerator Supplement to MES), the Canadian Institutes of Health Research
(New Investigator Salary Award to MES) and the Facult\'e de pharmacie at Universit\'e de Montr\'eal
(funding for AB and MES). 
{\it Conflict of Interest}: None declared.

\section{Appendix}
In the Appendix, we give the numerical results of the simulation study, the baseline  characteristics of our pregnancy data, the results of our application and the proof of the two lemmas.

\begin{table}[H]
\small\sf\centering
\caption{Simulation results (Data generating scenario 1). Estimates taken over 1000 generated datasets. $\widehat{\beta}_V$: average estimated value of the coefficients of the MSM, $\%Cov$: percent coverage of the selective confidence interval $\times 100$ (Standard CI for the linear model case), $\%sel$: percent selection of variables $\times 100$, FCR: False coverage rate $\times 100$, EM: T (variable is an effect-modifier) and F (variable is not an effect-modifier). The true values of the coefficients are $\beta_V =(0.5, 0, 1, 0)$}\label{algo}
\label{table:simu1}
\begin{tabular}{lllccccllccl }
\hline
& & \multicolumn{4}{c}{n=1000}& & &  \multicolumn{4}{c}{n=10000} \\

Coef & EM & {\it $\widehat{\beta}_V$ } &{\it  $\%sel$} & {\it $\%$Cov}& {\it FCR} & & & {\it $\widehat{\beta}_V  $ } & {\it  $\%sel$ } & {\it  $\%$Cov} & {\it FCR} \\
\hline

 & \\
   \multicolumn{12}{c}{(1)  \bf  $\bar{Q}$ \& $g$  model are correctly specified  } \\

{\bf $V_1$} & T &     0.46&    98&   96  &  \multirow{4}{*}{5}    &  & &     0.49 &  100   &  95 &     \multirow{4}{*}{6}\\
{\bf $V_2$} & F &     0.00&    21 &           &          &  & &     0.00  & 12    &   &          \\
{\bf $V_3$} & T &     0.98 &  100 & 95  &      &  & &        0.99&   100  &  95  &     \\
{\bf $V_4$} & F &    0.00&    21   &         &           &  & &            0.00  & 13   &     &         \\

 & \\
    \multicolumn{12}{c}{(2)  \bf  $\bar{Q}$ model  is correctly specified  } \\

{\bf $V_1$} & T &    0.46&   99&      96&   \multirow{4}{*}{6} &  & &        0.49&  100 &  95&    \multirow{4}{*}{6}\\
{\bf $V_2$} & F &    0.00&   21&               &    &  & &       0.00&  11  &            &  \\
{\bf $V_3$} & T &   0.98 &  100&       94&   &  & &         0.99&  100&   96&    \\
{\bf $V_4$} & F &   0.00&    19 &             &  &  & &       0.00&   12    &             &  \\

 & \\
    \multicolumn{12}{c}{(3)  \bf  $g$ model  is correctly specified  } \\

{\bf $V_1$} & T &  0.31&  55&      95&     \multirow{4}{*}{2}&  & &       0.47&   99 &  100&     \multirow{4}{*}{2} \\
{\bf $V_2$} & F &  0.01&  19&              &     &  & &       0.00&   14 &          &  \\
{\bf $V_3$} & T &  0.83&  92 &     100&   &  & &        0.99&   100&  99&      \\
{\bf $V_4$} & F & 0.00 & 26&                &    &  & &       0.00 &  22  &          &  \\

 & \\
   \multicolumn{12}{c}{(4)  \bf $\bar{Q}$ \& $g$ model are estimated using Hal  }  \\

{\bf $V_1$} & T   &    0.46  & 99    &  95 &   \multirow{4}{*}{ 6 }       &  & & 0.49    & 100  &    95  &   \multirow{4}{*}{6 }\\
{\bf $V_2$}  & F  &   0.00&   21  &          &          &  & &  0.00   &  12    &      &      \\
{\bf $V_3$}   & T &    0.98  &100 &      94   &     &  & &   1.00   &  100   &   95  &  \\
{\bf $V_4$}  & F &   0.00  & 22  &      &      &  & &      0.00&     13&          &     \\



& \\
   \multicolumn{12}{c}{(5)  \bf Naive Linear model  }  \\

{\bf $V_1$} & T &  0.69&   95 & 83 &  \multirow{4}{*}{19 }  & &  & 0.69 &  100 & 11 &  \multirow{4}{*}{ 65 }\\
{\bf $V_2$} & F &  0.15 &  12 & 88 &  &  & & 0.15&  67 & 33 &  \\
{\bf $V_3$} & T &  1.35 &   100  & 56 &  & &   & 1.36 &  100 & 0 &   \\
{\bf $V_4$} & F &  0.01 &  37 & 96 &  & &  &  0.00&  47 &  95 &   \\
& \\
   \multicolumn{12}{c}{(6)  \bf  Linear model correctly specified   }  \\

{\bf $V_1$} & T &  0.50 &   97 & 96   & \multirow{4}{*}{ 5}   & & &  0.50 &  100 & 95 &  \multirow{4}{*}{ 4}\\
{\bf $V_2$} & F &  0.00  &  6  & 94   &          & & & 0.00 &  5 & 95 &  \\
{\bf $V_3$} & T &  1.00 & 100  & 95    &         & & &  1.00 &  100 & 95 &   \\
{\bf $V_4$} & F & 0.00  &  4  & 96    &         & & & 0.00 &  4 &  96 &   \\

\hline
\end{tabular}
\end{table}

\begin{table}[H]
\small\sf\centering
\caption{Simulation results (Data generating scenario 2). Estimates taken over 1000 generated datasets. $\widehat{\beta}_V$: coefficients of the MSM, Cov: percent coverage of the selective confidence interval $\times 100$, $\%$sel: percent selection of variables $\times 100$, FCR: False coverage rate $\times 100$, EM: T (variable is an effect-modifier) and F (variable is not an effect-modifier). The true values of the coefficients are $\beta_V =(0.5, 0, 1, 0)$}\label{algo2}
\label{table:simu2}
\begin{tabular}{lllccccllccl }
\hline
& & \multicolumn{4}{c}{n=1000}& & &  \multicolumn{4}{c}{n=10000} \\

Coef & EM & {\it $\widehat{\beta}_V$ } &{\it  $\%sel$} & {\it $\%$Cov}& {\it FCR} & & & {\it $\widehat{\beta}_V  $ } & {\it  $\%sel$ } & {\it  $\%$Cov} & {\it FCR} \\
\hline

 & \\
   \multicolumn{12}{c}{(1)  \bf  $Q$ \& $g$  model are correctly specified  } \\

{\bf $V_1$} & T &   0.47  & 99  &     96&    \multirow{4}{*}{5} &  & &   0.49  &  100 &    95    &\multirow{4}{*}{5} \\
{\bf $V_2$} & F &      0.00&   20 &            &            &  & &   0.00 &    13 &                & \\
{\bf $V_3$} & T &     0.98&   100 &    95      &     &  & &     1.00  &  100 &      95 & \\
{\bf $V_4$} & F &   0.00&   23  &         &              &  & &   0.00   &  12   &           & \\

 & \\
    \multicolumn{12}{c}{(2)  \bf  $Q$ model  is correctly specified  } \\

{\bf $V_1$} & T &   0.47 &  99  &    97 &     \multirow{4}{*}{5}  &  & &       0.49 &   100&          94 &    \multirow{4}{*}{6}\\
{\bf $V_2$} & F &    0.00 &  20&                &         &  & &        0.00    & 11     &            &  \\
{\bf $V_3$} & T &    0.99 & 100&      95 &       &  & &        1.00      &  100 &         95&     \\
{\bf $V_4$} & F &    0.00 &  21. &                 &     &  & &       0.00       & 11    &               &  \\

 & \\
    \multicolumn{12}{c}{(3)  \bf  $g$ model  is correctly specified  } \\

{\bf $V_1$} & T &     0.32&   55&      99&         \multirow{4}{*}{2}&  & &     0.47&     99 &        99 &   \multirow{4}{*}{2}\\
{\bf $V_2$} & F &     0.01&   19&               &              &  & &    0.00&     14 &                 &\\
{\bf $V_3$} & T &     0.85 &  94 &     98  &            &  & &     0.99 &    100 &        99&    \\
{\bf $V_4$} & F &    -0.01&   24 &              &           &  & &     0.00   &  21     &            & \\

 & \\
   \multicolumn{12}{c}{(4)  \bf $Q$ \& $g$ model are estimated using Hal  }  \\

{\bf $V_1$} & T   &     0.47&   98&       97&    \multirow{4}{*}{5} &  & &  0.49&     100 &     95 &  \multirow{4}{*}{7}\\
{\bf $V_2$} & F   &      0.00&   22 &              &       &  & &    0.00   &  12     &               & \\
{\bf $V_3$} & T   &    0.98  & 100 &     94 &   &  & &      1.00     &100     & 95   & \\
{\bf $V_4$} & F   &     0.00&   22 &             &       &  & &      0.00   &  12    &           &   \\

& \\
   \multicolumn{12}{c}{(6)  \bf Linear model correctly specified   }  \\
{\bf $V_1$} & T &   0.50 &   89 & 96   & \multirow{4}{*}{5}    & & & 0.50 &  100 & 95 & \multirow{4}{*}{5}  \\
{\bf $V_2$} & F &  0.00 &  6 & 94    &          &  & & 0.00&  6 & 94 &  \\
{\bf $V_3$} & T &  1.00 & 100  & 94      &             & & & 1.00 &  100 & 95 &   \\
{\bf $V_4$} & F &  0.00 &  4 & 97    &          & &  & 0.00 &  4 &  96 &   \\

\hline
\end{tabular}
\end{table}

\begin{table}[H]
\small\sf\centering
\caption{Simulation results (Data generating scenario 3). Estimates taken over 1000 generated datasets. $\widehat{\beta}_V$: coefficients of the MSM, Cov: percent coverage of the selective confidence interval $\times 100$, $\%$sel: percent selection of variables $\times 100$, FCR: False coverage rate $\times 100$, EM: T (variable is an effect-modifier) and F (variable is not an effect-modifier). The true values of the coefficients are $\beta_V =(0.5, 0, 1, 0)$}\label{algo3}
\label{table:simu3}
\begin{tabular}{lllccccllccl }
\hline
& & \multicolumn{4}{c}{n=1000}& & &  \multicolumn{4}{c}{n=10000} \\

Coef & EM & {\it $\widehat{\beta}_V$ } &{\it  $\%sel$} & {\it $\%$Cov}& {\it FCR} & & & {\it $\widehat{\beta}_V  $ } & {\it  $\%sel$ } & {\it  $\%$Cov} & {\it FCR} \\
\hline

 & \\
   \multicolumn{12}{c}{(1)  \bf  $Q$ \& $g$  model are correctly specified  } \\

{\bf $V_1$} & T &    0.44&    94&        97&    \multirow{4}{*}{5}&  & &        0.49&  100&       96 &  \multirow{4}{*}{5} \\
{\bf $V_2$} & F &  0.00&    23  &                &          &  & &     0.00  &16    &            & \\
{\bf $V_3$} & T &    0.97&    100 &      95 &         &  & &      1.00  &100  &     97 &   \\
{\bf $V_4$} & F &   0.00 &   23  &                  &         &  & &    0.00 & 17   &           & \\

 & \\
    \multicolumn{12}{c}{(2)  \bf  $Q$ model  is correctly specified  } \\

{\bf $V_1$} & T &      0.45&   96&      97  &\multirow{4}{*}{6} &  & &    0.50 &  100 &     94  &\multirow{4}{*}{7} \\
{\bf $V_2$} & F &    0.00 &  20  &               &         &  & &  0.00&   13  &                &\\
{\bf $V_3$} & T &      0.98&   100 &    93&         &  & &     1.00&   100 &     95 &   \\
{\bf $V_4$} & F &    0.00 &   22   &               &       &  & &    0.00 &  12   &      &\\

 & \\
    \multicolumn{12}{c}{(3)  \bf   $g$ model  is correctly specified }  \\

{\bf $V_1$} & T &     0.34&   74 &     100&  \multirow{4}{*}{3} &  & &   0.49 &      100&   100&   \multirow{4}{*}{4} \\
{\bf $V_2$} & F &     0.01&   23&               &          &  & &   0.00 &      18  &               & \\
{\bf $V_3$} & T &    0.91 &  99 &     97 &          &  & &     0.99&      100&    96 &   \\
{\bf $V_4$} & F &   0.00&   25 &              &           &  & &   0.00 &      24 & &\\

 & \\
   \multicolumn{12}{c}{(4)  \bf $Q$ \& $g$ model are estimated using Hal  }  \\

{\bf $V_1$} & T &      0.45&   95&      95&  \multirow{4}{*}{6}&  & &    0.49 &    100&      95&    \multirow{4}{*}{5}\\
{\bf $V_2$} & F &     0.00&   24 &               &          &  & &   0.00  &   16   &             &\\
{\bf $V_3$} & T &      0.98  & 100 &     94&          &  & &    1.00 &    100&      96 &   \\
{\bf $V_4$} & F &      0.00&   23 &               &         &  & & 0.00 &    16   &      &\\

& \\
   \multicolumn{12}{c}{(5)  \bf Naive Linear model    }  \\
{\bf $V_1$} & T &  0.60 &   89 & 93  & \multirow{4}{*}{10}     & & & 0.59 &  100 & 63 & \multirow{4}{*}{43}\\
{\bf $V_2$} & F &  0.10 &  76 & 92    &          &  & & 0.10&  35 & 65 &  \\
{\bf $V_3$} & T &  1.21 & 100  & 81    &             & & & 1.21 &  100 & 58 &   \\
{\bf $V_4$} & F &  0.01 &  38 & 96   &          & &  &-0.00&  44 &  96 &   \\

& \\
   \multicolumn{12}{c}{(6)  \bf Linear model correctly specified   }  \\
{\bf $V_1$} & T &  0.50 &  98 & 96   & \multirow{4}{*}{5}    & & & 0.50 &  100 & 95 & \multirow{4}{*}{5}\\
{\bf $V_2$} & F &  0.00 &  4 & 96    &          &  & & 0.00&  5 & 95 &  \\
{\bf $V_3$} & T &  1.00 & 100   & 95    &          & & & 1.00 &  100 & 95 &   \\
{\bf $V_4$} & F &  0.00 &  5 & 95    &          & &  &  0.00&  5 &  95 &   \\

\hline
\end{tabular}
\end{table}

\begin{table}[H]
\small\sf\centering
\caption{Simulation results for smaller sample size ($n=100$). Estimates taken over $500$ generated datasets. $\widehat{\beta}_V$: coefficients of the MSM, Cov: percent coverage of the selective confidence interval $\times 100$, $\%$sel: percent selection of variables $\times 100$, FCR: False coverage rate $\times 100$, EM: T (variable is an effect-modifier) and F (variable is not an effect-modifier). The true values of the coefficients are $\beta_V =(0.5, 0, 1, 0)$}\label{ssim_small}
\label{table:smallsize}
\begin{tabular}{lllccllccllcccl }
\hline
& & \multicolumn{4}{c}{scenario 1} &  \multicolumn{4}{c}{scenario 2} &  \multicolumn{4}{c}{scenario 3}\\

Coef & EM & {\it $\widehat{\beta}_V$ } &{\it  $\%sel$} & {\it Cov}& {\it FCR}  & {\it $\widehat{\beta}_V  $ } & {\it  $\%sel$ } & {\it  Cov} & {\it FCR} &  {\it $\widehat{\beta}_V  $ } & {\it  $\%sel$ } & {\it  Cov} & {\it FCR}\\
\hline

& \\
 & & 
   \multicolumn{12}{c}{(1)  \bf  $Q$ \& $g$  model are correctly specified  } \\

{\bf $V_1$} & T &    0.39&    52&        87&    \multirow{4}{*}{8}&           0.34&  49&       88 &  \multirow{4}{*}{9} &            0.30 &  41&       89 &  \multirow{4}{*}{10}\\
{\bf $V_2$} & F &  -0.01&    22  &                &          &        -0.01  &25  &            &  &        0.02 &24  &            &  \\
{\bf $V_3$} & T &     0.85&    86 &      94 &         &        0.78  &80  &     96 &   &       0.78  &71  &     93 &  \\
{\bf $V_4$} & F &   0.01 &  28.  &                  &         &       0.00 & 25  &           & &       0.00 & 24  &           & \\

 & \\
 & &   \multicolumn{12}{c}{(2)  \bf  $Q$  model is correctly specified  } \\

{\bf $V_1$} & T &    0.38&    53&       91&    \multirow{4}{*}{7}&          0.36&  50&      88 &  \multirow{4}{*}{8} &            0.29 &  41&       89 &  \multirow{4}{*}{10} \\
{\bf $V_2$} & F &  -0.03&    27 &                &          &   0.00  &21    &            & &        0.01 &20  &            & \\
{\bf $V_3$} & T &     0.83&    85 &      98 &         &      0.79  &8  &     97 &  &       0.76  &72  &     93 & \\
{\bf $V_4$} & F &   -0.02 &  25  &                  &         &      0.00 & 27  &           & &       0.00 & 21  &           &  \\

& \\
 & &   \multicolumn{12}{c}{(3)  \bf  $g$  model is correctly specified  } \\

{\bf $V_1$} & T &    0.24&    20&        97&    \multirow{4}{*}{9}&        0.24&  25&       98 &  \multirow{4}{*}{6} &             0.26 &  25&       91 &  \multirow{4}{*}{9}\\
{\bf $V_2$} & F &  0.04&    16  &                &          &       0.04  &1    &            & &        0.04&26 &            &  \\
{\bf $V_3$} & T &     0.51&    29 &      90 &         &      0.59  &45  &     95 &  &       0.68 &47 &     88 & \\
{\bf $V_4$} & F &   0.01 &  21  &                  &         &     0.02 & 23   &           & &       0.00 & 25  &           &  \\

& \\
& &  \multicolumn{12}{c}{(4)  \bf $Q$ \& $g$  model are  estimated using Hal  } \\

{\bf $V_1$} & T &    0.39&    54&        83&    \multirow{4}{*}{10}&        0.36&  51&      85 &  \multirow{4}{*}{9}  &             0.32 &  45&       79 &  \multirow{4}{*}{11}\\
{\bf $V_2$} & F &  0.00&    30  &                &          &      0.01  &27    &            &  &        0.00 &27  &            &  \\
{\bf $V_3$} & T &     0.84&    87 &      96 &         &     0.79  &81  &     96 &  &       0.80  &82  &     95 & \\
{\bf $V_4$} & F &  0.00 &  27  &                  &         &    0.01 & 27   &           & &       -0.02 & 24  &           &  \\

 \hline
\end{tabular}
\end{table}

\begin{table}[H]
\small\sf\centering
\caption{Simulation results (Data generating scenario 1 with $50$ noise covariates). Estimates taken over 100 generated datasets. $\widehat{\boldsymbol\beta}_V$: coefficients of the MSM, Cov: percent coverage of the selective confidence interval, $\%$sel: percent selection of variables, FCR: False coverage rate, EM: T (variable is an effect-modifier) and F (variable is not an effect-modifier). The true values of the coefficients are $\beta_V =(0.5, 0, 1, 0,...,0)$}\label{table:HD}
\begin{tabular}{lllccccllccl }
\hline
& & \multicolumn{4}{c}{n=1000}& & &  \multicolumn{4}{c}{n=10000} \\

Coef & EM & {\it $\widehat{\beta}_V$ } &{\it  $\%sel$} & {\it $\%$Cov}& {\it FCR} & & & {\it $\widehat{\beta}_V  $ } & {\it  $\%sel$ } & {\it  $\%$Cov} & {\it FCR} \\
\hline

 & \\
   \multicolumn{12}{c}{(1)  \bf  Estimates related to the potential EM that are not noise covariates.} \\

{\bf $V_1$} & T &   0.43  & 100  &     100&    \multirow{4}{*}{15}      &  & &     0.48  & 100    &   100    &\multirow{4}{*}{15} \\
{\bf $V_2$} & F &      0.00&   14 &             &            &  & &     0.00 & 15    &                & \\
{\bf $V_3$} & T &     0.95 &   100 &    91      &       &  & &      0.99 & 100   &   90    & \\
{\bf $V_4$} & F &   0.01&   15   &          &                  &  & &      0.00 & 14      &           & \\

& \\
   \multicolumn{12}{c}{(2)  \bf  Summary of the 50 potential EM that are noise covariates. } \\

{\bf min} &  &  -0.01  & 7.0  &     &                        &  & &     0.00 & 5  &       & \\
{\bf $Q_1$} &  &      0.00&   12 &             &            &  & &     0.00  & 11     &                & \\
{\bf median} &  &     0.00 &   14 &          &              &  & &      0.00  & 13   &       & \\
{\bf $Q_3$} &  &   0.00&   16   &          &                      &  & &     0.00  &15      &           & \\
{\bf max}&  &   0.01&   23  &          &                           &  & &     0.00 & 22      &           & \\

\hline
\end{tabular}
\end{table}

\begin{table}[H]
\small\sf\centering
\caption{Baseline Characteristics of mothers in the cohort extraction ($N=4,707$) . }  
\label{table:Baseline}
\begin{tabular}{l c c } 
\hline
\textbf{} & \textbf{No ICS}& \textbf{ICS} \\ [0.1ex]  
\textbf{Characteristics} & \textbf{N ($\%$)}  & \textbf{ N ($\%$)} \\
\hline 

Cohort size & 2272 (100) &  2435 (100) \\

Age & & \\
~~~ $< 18$ & 45 (1.9)  &   60 (2.4) \\
~~~  18-34 & 1958 (86.1)    & 2041 (83.8) \\
~~~  $>34$ &  269 (11.8)  &    334(13.7) \\
Sex of the newborn &  1149 (51.0)  & 1271 (52.0)   \\
Welfare recipient  &  1126 (50.0)  & 1429 (59.0)   \\
Urban residence  &   476 (18.0)  & 407 (20.0) \\
Hypertension & 61 (3.0) &  83 (3.0) \\
Diabetes & 73 (3.0) & 81 (3.0) \\
COPD &  28 (1.0) &  56 (2.0) \\
Cyanotic heart disease & 7 (0.0) & 8 (0.0 )\\

Antiphospholipid syndrome &  12 (1.0) & 13 (1.0) \\
Uterine disorder &  264 (12.0) &   331 (14.0) \\
Epilepsy &  18 (1.0) &   23 (1.0) \\
Obesity &  87 (4.0) &   127 (5.0) \\
Lupus & 1 (0.0) & 2 (0.0) \\
Collagenous vascular disease & 6 (0.0) & 6 (0.0) \\
Cushing's syndrome & 4 (0.0) & 4 (0.0) \\
Oral corticosteroids one year before pregnancy & 234 (10.0) &  281(12.0) \\
Oral SABA use one year before pregnancy &  16 (1.0) &  8 (0.0) \\
At least one dose of inhaled SABA taken per week  &  1523 (67.0) &  1332 (55.0) \\
HIV & 3 (0.0) &  1 (0.0) \\
Cytomegalovirus infection & 3 (0.0)  & 12 (0.0) \\
Leukoteriene-receptor antagonists & 33 (1.0) & 30 (1.0) \\
Theophylline use one year before pregnancy & 0 (0.0)  & 0 (0.0) \\
Intranasal corticosteroids &  243 (11.0) &  318 (13.0) \\
Folic acid one year before pregnancy & 18 (1.0) &  43 (2.0) \\
Teratogenes taken one year before  &  0 (0.0) & 0 (0.0) \\
Medication for epilepsy one year before pregnancy & 29 (1.0) & 48 (2.0) \\
Warfarin one year before pregnancy & 7(0.0) & 10 (0.0) \\
Use of beta-bloqueur one year before pregnancy &  19 (1.0) & 26 (1.0)\\
Asthma exacerbation one year before pregnancy & 377 (17.0) & 411 (17.0) \\
hospitalization for asthma & 1079 (47.0) & 809 (33.0) \\
Chromosomal anomalies &  6 (0.0) & 4 (0.0) \\
Cumulative dose of ICS in days (mean (SD)) & 51.6 (72.8)  &54.0 (85.8) \\
One year cumulative dose of ICS before pregnancy (mean (SD)) & 151 (32.0)  & 101.5 (126.3) \\
At least one emergency department visit for asthma &  260 (7.0) & 265 (19.0) \\
At least one hospitalization for asthma  & 5 (0.0) &  8 (1.0) \\

\hline 
\end{tabular}
\label{table:char} 
\end{table}

\begin{table}[H]
\small\sf\centering
\caption{Estimates of the coefficients associated with interaction terms using naive linear model ($n=4707$).} 
\label{table:LinearAppli} 
\begin{tabular}{l c c c } 
\hline
\textbf{Variables} & \textbf{Estimate ($\widehat{\beta}_j$)} & \textbf{ STD} & \textbf{P-value}\\ [0.5ex] 
\hline

Intercept         &    3.153 & & \\
CS:At least one dose of inhaled SABA taken per week    &     -0.002   &0.039   & 0.940\\   
CS:Leukoteriene-receptor antagonists       &      -0.365   &0.142   & 0.010*\\  
CS:Intranasal corticosteroids        &       0.063&  0.051 &0.214\\    
CS:Folic acid one year before pregnancy             &     -0.129  & 0.159   & 0.415\\  
CS:Medication for epilepsie   &      -0.136 &  0.135   & 0.313\\   
CS:Warfarin    &     -0.386 &  0.277  & 0.164\\  
CS:Beta-blockers   &      -0.287 & 0.173 & 0.097\\  
CS:Asthma exacerbation     &           0.062   & 0.069   & 0.368\\   
CS:At least one hospitalization for asthma       &        0.017  & 0,036   & 0.624\\    
CS:At least one emergency department
visit for asthma         &         0.067 &  0.055   & 0.223\\   
CS:COPD     &             0.141 &  0.130    & 0.280\\   
CS:Cyanotic heart disease   &         -0.345&  0.292   & 0.237\\   
CS:Oral corticosteroids one year before   &           -0.081  & 0.081 & 0.319\\   
CS:Obesity        &      0.053 &  0.080   & 0.508\\   
CS:Uterine disorder        &        -0.036 &  0.050  & 0.460\\    
CS:Oral SABA use one year before   &         -0.025  & 0.244   & 0.918\\  
CS:Antiphospholipid syndrome  &         0.394  & 0.227 &   0.083\\  
CS:Sex of new born       &       -0.031 &  0.032 & 0.335\\    
CS:Welfare recipient          &        -0.043  & 0.033  & 0.1871\\
CS:Rural/non-rural residence indicator    &          0.021 & 0.042  &   0.602\\    
CS:Hypertension           &        0.028   &  0.098  & 0.774\\  
CS:Diabetes          &         -0.105   &0.092  &  0.255\\   
CS:Chromosomal anomalies   &          -1.230 &  0.361 &  0.0006*\\
CS:Cytomegalovirus infection    &          0.146 &  0.360 &0.683\\
\hline 
\end{tabular}
\end{table}

\begin{table}[H]
\small\sf\centering
\caption{Estimates of the selected MSM coefficients using adaptive lasso ($n=4707$) with $95\%$ Post selection  interval for the selected variables. *: means significant variables} 
\label{table:MSMappli} 
\begin{tabular}{l c c c } 
\hline
\textbf{Variables} & \textbf{Estimate ($\widehat{\beta}_j$)} & \textbf{ CI Low} & \textbf{CI up}\\ [0.5ex] 
\hline 

   \multicolumn{4}{c}{  \bf High adaptive LASSO for $Q$ \& $g$   }  \\
Intercept        &  0.018 &  &  \\   
Leukoteriene-receptor antagonists*        &  -0.177& -0.502 & -0.031 \\ 
Warfarin    &  -0.146& -0.745 &  0.311 \\
Chromosomal anomalies*       &  -0.777& -1.420 & -0.285 \\

\hline 
\end{tabular}
\end{table}

\textbf{Proof of Lemma 1}:
Denote $\bar{Q}_n$ (respectively $g_n$) an estimator of $\bar{Q}$ (respectively $g$). We have:

\begin{multline*}
    E_{P_0}( D(\bar{Q}_n,g_n)|\boldsymbol V) = E_{P_0}\left\{ \displaystyle\frac{2A-1}{g_n(A|\boldsymbol W)}(Y-\bar{Q}_n(A,\boldsymbol W))+\bar{Q}_n(1,\boldsymbol W)-\bar{Q}_n(0,\boldsymbol W) |\boldsymbol V\right\} \\
    =  E_{P_0}\left\{ \displaystyle\frac{2A-1}{g_n(A|\boldsymbol W)}(Y-\bar{Q}_n(A,\boldsymbol W )|\boldsymbol V\right\}+E_{P_0}(\bar{Q}_n(1,\boldsymbol W)-\bar{Q}_n(0,\boldsymbol W)|\boldsymbol V) + \psi_0(\boldsymbol V)- \psi_0(\boldsymbol V)\\
= \psi_0(\boldsymbol V) +  E_{P_0}\left\{ [\bar{Q}_n(1,\boldsymbol W)-\bar{Q}_n(0,\boldsymbol W)]- [\bar{Q}_0(1,\boldsymbol W)-\bar{Q}_0(0,\boldsymbol W)]  |\boldsymbol V\right\} +\\E_{P_0}\left\{ \displaystyle\frac{2A-1}{g_n(A|\boldsymbol W)}(Y-\bar{Q}_n(A,\boldsymbol W))|\boldsymbol V \right\} \\
=  \psi_0(\boldsymbol V) +\int_{\boldsymbol W}( [\bar{Q}_n(1,\boldsymbol W)-\bar{Q}_n(0,\boldsymbol W)] - [\bar{Q}_0(1,\boldsymbol W)-\bar{Q}_0(0,\boldsymbol W)]  +\\ \displaystyle \frac{P_0(1|\boldsymbol W)}{g_n(1|\boldsymbol W)} \left\{ \bar{Q}_0(1,\boldsymbol W)- \bar{Q}_n(1,\boldsymbol W) \right\}- \displaystyle \frac{P_0(0|\boldsymbol W)}{g_n(0|\boldsymbol W)} \left\{ \bar{Q}_0(0,\boldsymbol W)- \bar{Q}_n(0,\boldsymbol W) \right\} ) d P_0(\boldsymbol W|\boldsymbol V)  \\
=  \psi_0(\boldsymbol V) + \int_{\boldsymbol W} [\displaystyle \frac{P_0(1|\boldsymbol W)}{g_n(1|\boldsymbol W)}-1)(\bar{Q}_0(1,\boldsymbol W)-\bar{Q}_n(1,\boldsymbol W)) +\\ (\displaystyle \frac{P_0(0|\boldsymbol W)}{g_n(0|\boldsymbol W)}-1)(\bar{Q}_0(0,\boldsymbol W)-\bar{Q}_n(0,\boldsymbol W)) ]d P_0(\boldsymbol W|\boldsymbol V) 
\end{multline*}

 Then $E_{P_0}( D(\bar{Q}_n,g_n)|\boldsymbol V)  \rightarrow  \psi_0(\boldsymbol V)$ if $g_n(A|\boldsymbol W)$ or $\bar{Q}_n(A,\boldsymbol W)$ is consistently estimated.
 \\

\textbf{Proof of Lemma 2}: Let $D=D(\bar{Q}_0,g_0)$ represent the oracle pseudo function.
our method minimizes the expected risk function below with respect to $\beta$:
$$  \{ (D- \sum_jV^{(j)}\beta_j)^2 +\lambda\sum_{j=1}^{p}\widehat{w}_j |\beta_j| \}$$ where $\widehat{\omega}_j = 1 / |\tilde{\beta}_j|^{\gamma}$, $j=1,...,p$, for some $\gamma > 0 $.\\ Let $\epsilon = D-\sum_j V^{(j)}\beta_j$ be the residual of the penalized linear regression of the oracle pseudo function $D$ on $\boldsymbol V$.
We have $\epsilon^T\boldsymbol V/\sqrt{n}  \overset{d}{\rightarrow} N(0,\sigma^2C)$. The rest of the proof in \cite{R17} follows. We assume $\frac{1}{n}\boldsymbol V^T\boldsymbol V \rightarrow C$ with C a positive definite matrix.  \\
 Here, as $D=D(\bar{Q}_0,g_0)$ represents the true pseudo function, we conjecture the result will also hold if both $(Q_0, g_0)$ are consistently estimated at fast enough rates.
 \\

\textbf{Asymptotic Normality (Follows exactly the proof in \cite{R17})} \\
Let $\widehat{\beta} =\beta +\frac{\mu}{\sqrt{n}}$ and  $\phi_n(\mu)=|| D -  \sum_{j=1}^{p}V^{(j)}(\beta_j +\frac{\mu_j}{\sqrt{n}} )||^2 + \lambda \sum_{j=1}^{p}\widehat{w}_j |\beta_j+\frac{\mu_j}{\sqrt{n}} |$. We have $\phi_n(\mu)-\phi_n(0) = V_4^{(n)} (\mu)$  with
$$ V_4^{(n)} (\mu) =\underbrace{ \mu^T(\frac{1}{n}\boldsymbol V^T\boldsymbol V)\mu }_{ A} - \underbrace{2\frac{\epsilon^T\boldsymbol V}{\sqrt{n}} \mu   }_{ B} + \underbrace{\frac{\lambda}{\sqrt{n}} \sum_{j=1}^{p}\widehat{w}_j\sqrt{n} (  |\beta_j +\frac{\mu_j}{\sqrt{n}} | - |\beta_j|) }_{C} $$
We know that B has a normal distribution with mean zero and covariance matrix. It can be shown \cite{R17} that $C  \underset{p}{\rightarrow} 0$. Following \cite{R17} who used the result of Knight and Fu (2006), we have $\widehat{\mu}_{\mathcal{A}}   \underset{d}{\rightarrow} C^{-1}_{11}\mathcal{W}_{\mathcal{A}}$  and  $\mathcal{W}_{\mathcal{A}} = N(0,\sigma^2 C^{-1}_{11})$ which illustrate the asymptotic normality. $\mathcal{A}$ is the set of true coefficient in the model.
\\
\textbf{Consistency (Follows almost exactly the proof in \cite{R17})} \\
Let $\mathcal{A}$  be the set of true variable in the model and $\mathcal{A}_n^*$ be the selected set of variables. Assuming that $j' \not\in \mathcal{A}$, it suffices to show that $P( j' \in \mathcal{A}_n^* ) \rightarrow 0 $. \\Suppose $\lambda n^{(-1/2+\gamma/2)} \rightarrow \infty$.
If $ j' \in \mathcal{A}_n^* $, that means, if the variable associated with $j'$ would belong to the true set of variables, then $\widehat{\mu}_{j'}$ would be solution of minimizing the function $\phi_n(\mu)$ with respect to $\mu$. Therefore, $\widehat{\beta}_{j'}=\beta +\frac{\widehat{\mu}_{j'}}{\sqrt{n}}$. By the Karush-Kuhn-Tucker (KKT) conditions, we can write $ 2V^{(j')^T}(D-\boldsymbol V\widehat{\beta})=\lambda \widehat{w}_{j'}$\\
Note that:
$$\frac{\lambda \widehat{w}_{j'}}{\sqrt{n}} = \frac{\lambda}{\sqrt{n}}n^{\gamma/2} \frac{1}{|\sqrt{n}\tilde{\beta}_{j'}|^{\gamma}}   \underset{p}{\rightarrow} \infty$$ 
and
$$\frac{2V^{(j')^T}(D-\boldsymbol V\widehat{\beta})}{\sqrt{n}} =2 \frac{V^{(j')^T}\boldsymbol V\sqrt{n}(\beta- \widehat{\beta})}{n} +2\frac{V^{(j')^T}\epsilon}{\sqrt{n}}$$
The first term in the right side of the last equation converges to a normal distribution with mean zero (asymptotic normality) and the second part converge to $N(0, 4||V^{(j')}||^2\sigma^2)$.  By Slutsky's theorem, $2V^{(j')^T}(D-\boldsymbol V\widehat{\beta})/\sqrt{n}$ follows a normal distribution with mean zero. Thus,
$$P(j' \in \mathcal{A}_n^* )  \leq  P(2V^{(j')^T}(D-\boldsymbol V\widehat{\beta})=\lambda \widehat{w}_{j'}) \rightarrow 0 $$

We then show that, if $j' \not\in \mathcal{A}$, $P(j' \in \mathcal{A}_n^* ) \rightarrow 0 $.

\bibliographystyle{DeGruyter}

\end{document}